\documentclass{ws-mpla}
\usepackage{amstext}
\usepackage{graphicx}
\usepackage{amssymb}

\begin{document}

\title{Decoherence in quantum cosmology and the cosmological constant }

\author{Torsten Asselmeyer-Maluga }

\address{German Aerospace Center, Berlin and Loyola University, New Orleans, LA, USA \\ torsten.asselmeyer-maluga@dlr.de}

\author{Jerzy Kr\'ol}

\address{University of Silesia, Institute of Physics, ul. Uniwesytecka 4,
40-007 Katowice, Poland \\ iriking@wp.pl}

\maketitle

\pub{Received (Day Month Year)}{Revised (Day Month Year)}
\begin{abstract}
We discuss a spacetime having the topology of $S^{3}\times\mathbb{R}$
but with a different smoothness structure. The initial state of the
cosmos in our model is identified with a wildly embedded 3-sphere
(or a fractal space). In previous work we showed that a wild embedding
is obtained by a quantization of a usual (or tame) embedding. Then
a wild embedding can be identified with a (geometrical) quantum state.
During a decoherence process this wild 3-sphere is changed to a homology
3-sphere. We are able to calculate the decoherence time for this process.
After the formation of the homology 3-sphere, we obtain a spacetime
with an accelerated expansion enforced by a cosmological constant.
The calculation of this cosmological constant gives a qualitative
agreement with the current measured value.
\keywords{decoherence in cosmology; homology 3-spheres; cosmological constant.}
\end{abstract}

\ccode{PACS Nos.:04.60.Gw, 02.40.Ma, 04.60.Rt}

\section{Introduction}
General relativity (GR) has changed our understanding of spacetime.
In parallel, the appearance of quantum field theory (QFT) has modified
our view of particles, fields and the measurement process. The usual
approach for the unification of QFT and GR, to a quantum gravity,
starts with a proposal to quantize GR and its underlying structure,
spacetime. There is a unique opinion in the community about the relation
between geometry and quantum theory: The geometry as used in GR is
classical and should emerge from a quantum gravity in the limit (Planck's
constant tends to zero). Most theories went a step further and try
to get a spacetime from quantum theory. Then, the model of a smooth
manifold is not suitable to describe quantum gravity. But, there is
no sign for a discrete spacetime structure or higher dimensions in
current experiments. Hence, quantum gravity based on the concept of
a smooth manifold should also able to explain the current problems
in the standard cosmological model ($\Lambda$CDM) like the appearance
of dark energy/matter, or the correct form of inflation etc. But before
we are going in this direction we will motivate the usage of the 'good
old' smooth manifold as our basic concept.

When Einstein developed GR, his opinion about the importance of general
covariance changed over the years. In 1914, he wrote a joint paper
with Grossmann. There, he rejected general covariance by the now famous
hole argument. But after a painful year, he again considered general
covariance now with the insight that there is no meaning in referring
to the \emph{spacetime point A} or the \emph{event A}, without further
specifications. Therefore the measurement of a point without a detailed
specification of the whole measurement process is meaningless in GR.
The reason is simply the diffeomorphism-invariance of GR which has
tremendous consequences. Furthermore, GR do not depend on the concrete
topology of spacetime. All restrictions on the topology of the spacetime
were formulated using additional physical conditions like causality
(see \cite{HawEll:94}). This ambiguity increases in the 80's when
the first examples of exotic smoothness structures in dimension 4
were found. The (smooth) atlas of a smooth 4-manifold $M$ is called
the smoothness structure (unique up to diffeomorphisms). One would
expect that there is only one smooth atlas for $M$, all other possibilities
can be transformed into each other by a diffeomorphism. But in contrast,
the deep results of Freedman \cite{Fre:82} on the topology of 4-manifolds
combined with Donaldson's work \cite{Don:83} gave the first examples
of non-diffeomorphic smoothness structures on 4-manifolds including
the well-known $\mathbb{R}^{4}$. Much of the motivation can be found
in the FQXI essay \cite{AsselmeyerFQXI-Essay2012}. Here we will discuss
another property of the exotic smoothness structure: its quantum geometry.

\section{Exotic $S^3\times {\mathbb R}$}
Let us consider the spacetime with topology $S^{3}\times\mathbb{R}$
where the 3-sphere has growing radius. This spacetime can admit uncountable
many, different (=non-diffeomorphic) smoothness structures, denoted
by $S^{3}\times_{\theta}\mathbb{R}$ (a first example was constructed
in \cite{Fre:79}). For the construction of $S^{3}\times_{\theta}\mathbb{R}$,
one needs a homology 3-sphere $\Sigma$, i.e. a compact 3-manifold
with the same homology like the 3-sphere. The Poincare sphere (or
the dodecahedral space used in cosmology, see \cite{dodecaeder:03})
is an example. Usually, every homology 3-sphere is the boundary of
a contractable 4-manifold but not every homology 3-sphere is the boundary
of a SMOOTH, contractable 4-manifold. We used this fact to formulate
restrictions on possible smooth spacetimes in cosmology \cite{AsselmKrol-2012i}.
For the construction of an exotic $S^{3}\times_{\theta}\mathbb{R}$
one needs the following pieces:
\begin{enumerate}
\item $W_{1}$ as cobordism between $\Sigma$ and its one-point complement
$\Sigma\setminus pt.$ and
\item $W_{2}$ as cobordism between $\Sigma\setminus pt.$ and $\Sigma\setminus pt.$. 
\end{enumerate}
Then the non-compact 4-manifold 
\begin{equation}
W=\ldots\cup-W_{2}\cup-W_{2}\cup\left(-W_{1}\cup W_{1}\right)\cup W_{2}\cup W_{2}\cup\ldots\label{eq:exotic-S3xR}
\end{equation}
 (see \cite{Fre:79}, $-W_{i}$ has reversed orientation ) is homeomorphic
to $S^{3}\times\mathbb{R}$ but not diffeomorphic to it, i.e. $W=S^{3}\times_{\theta}\mathbb{R}$.
But $S^{3}\times_{\theta}\mathbb{R}$ has the topology of $S^{3}\times\mathbb{R}$,
i.e. for every $t\in\mathbb{R}$ there is a 3-sphere $S^{3}\times\left\{ t\right\} $
topologically but not smoothly embedded into $S^{3}\times_{\theta}\mathbb{R}$.
Or,\\
\emph{The 3-sphere $S^{3}\hookrightarrow S^{3}\times_{\theta}\mathbb{R}$
is wildly embedded, i.e. it is only represented (better triangulated)
by infinite many polyhedrons. The wild 3-sphere is a fractal space.}
\\
In \cite{AsselmeyerKrol2011b,AsselmeyerKrol2013} we discussed wild
embeddings and its relation to quantum geometry. We were able to show
that the (deformation) quantization of a tame embedding (see the appendix
for a definition) is a wild embedding. The idea of the proof can be
simply expressed in the formalism of GR. First we consider a tame
embedding $S^{3}\hookrightarrow S^{3}\times\mathbb{R}$ for a non-exotic
spacetime. The 3-sphere can be triangulated and we consider the 1-skeleton,
i.e. a finite graph. The holonomies along this graph with respect
to a suitable connection representing the geometry. The geometry of
$S^{3}$ can be at least locally approximated by a homogenous ($SO(3)$
invariant) metric (in the Cartan geometry formalism, see \cite{Wise2009,Wise2010}).
The observables in this theory (as the functions over the space of
holonomies) form a Poisson algebra \cite{Goldman1984}. The (deformation)
quantization of this Poisson algebra \cite{Turaev1991} transformed
the graph into a knotted, infinite graph. As shown by us, this knotted
graph is a wild embedding which can be interpreted as the quantum
state. This quantum state contains an infinite number of homogenous
metrics, i.e. metrics with different values of the curvature. Then
the transition of a wild embedding to a tame embedding is the transition
from the quantum state to a classical state which can be interpreted
as decoherence. 

\section{Decoherence as Topology-change}
\begin{figure}
\includegraphics[scale=0.5]{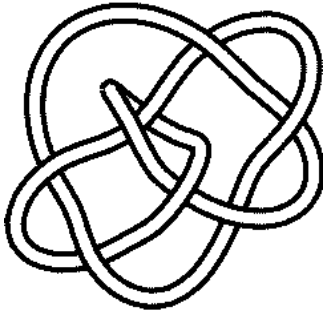}

\caption{Knot $8_{10}${\label{fig:Knot-8_10}}}
\end{figure}

Now we will interpret the wild embedded 3-sphere in cosmology. In
our model $W$ of the exotic $S^{3}\times_{\theta}\mathbb{R}$ we
made a rescaling so that the 3-sphere at $t=-\infty$ is the initial
state of the cosmos (at the big bang), i.e. we assume that the cosmos
starts as a small 3-sphere (of Planck radius). As we claimed above,
this 3-phere is a wildly embedded 3-sphere $S_{\theta}^{3}$. In the
model (\ref{eq:exotic-S3xR}) of $W$ above we have the part $-W_{1}\cup W_{1}$
as a cobordism from $\Sigma\setminus pt.$ to $\Sigma$ and back.
As Freedman \cite{Fre:79} showed this cobordism is equivalent to
a cobordism from $S^{3}\setminus pt.$ to $\Sigma$ and back. But
now we can identify the wild $S_{\theta}^{3}$ (partly) with $S^{3}\setminus pt.$
in this cobordism. But then we obtain a transition from the wild $S_{\theta}^{3}$
(the quantum state) to the homology 3-sphere $\Sigma$, which is smoothly
embedded $\Sigma\hookrightarrow W=S^{3}\times_{\theta}\mathbb{R}$.
Or,
\begin{equation}
\mbox{quantum state}\quad S_{\theta}^{3}\stackrel{\mbox{decoherence}}{\longrightarrow\longrightarrow\longrightarrow}\mbox{classical state}\quad\Sigma\label{eq:decoherence}
\end{equation}
and we studied this process in \cite{AsselmKrol-2013b} more carefully.
In case of a hyperbolic homology 3-sphere, we showed that there is
a single parameter, a topological invariant 
\[
\vartheta=\frac{3\cdot vol(\Sigma)}{2\cdot CS(\Sigma)}
\]
of $\Sigma$, which characterizes all properties of this process (\ref{eq:decoherence}).
Hyperbolic $n-$manifolds for $n>2$ show Mostow rigidity, i.e. every
diffeomorphism or conformal transformation is an isometry. Therefore
the (unit) volume $vol(\Sigma)$ and the Chern-Simons functional (as
integral over the scalar curvature of $\Sigma$) are topological invariants.
In \cite{AsselmKrol-2013b}, we interpreted the process (\ref{eq:decoherence})
as inflation and calculated the exponential increase to be
\begin{equation}
\exp\left(\frac{3\cdot vol(\Sigma)}{2\cdot CS(\Sigma)}\right)=\exp(\vartheta)\label{eq:expansion-rate}
\end{equation}
Now we further interpret the process as decoherence of a quantum state
(the wild 3-sphere) to the classical state (the tame hyperbolic homology
3-sphere). But then the decoherence time can be calculated from \cite{AsselmKrol-2013b}.
Assume the Planck time $T_{Planck}$ as time scale for the quantum
state then we obtain for the decoherence time $T_{decoherence}$
\[
T_{decoherence}=T_{Planck}\cdot\sum_{n=0}^{5}\frac{\vartheta^{n}}{n!}
\]
(see section 4.3 in \cite{AsselmKrol-2013b}). Consider as an example
the hyperbolic homology 3-sphere $\Sigma(8_{10})$ obtained by Dehn
surgery along the knot $8_{10}$ (in Rolfson notation), see Fig. \ref{fig:Knot-8_10}.
 Then we obtain the values (using SnapPea of J. Weeks)
\begin{eqnarray*}
vol(\Sigma(8_{10})) & = & 8.65115...\\
CS(\Sigma(8_{10})) & = & 0.15616...
\end{eqnarray*}
or 
\[
\vartheta=\frac{3\cdot vol(\Sigma)}{2\cdot CS(\Sigma)}\approx 83.131....
\]
and the decoherence time 
\[
T_{decoherence}=T_{Planck}\cdot\sum_{n=0}^{5}\frac{\vartheta^{n}}{n!}\approx3\cdot10^{-36}s\quad.
\]
According to this model, the cosmos starts in a quantum state (represented
by a wild 3-sphere of Planck size) which undergoes a transition to
a homology 3-sphere (classical state) by a decoherence process with
time $\approx 3\cdot 10^{-36}s$. 

\section{The Cosmological Constant}
But what does the model predict for
the future evolution? After the formation of the (hyperbolic)
homology 3-sphere, the whole spacetime admits a hyperbolic structure
(metric of negative scalar curvature). At the same time, the hyperbolic
homology 3-sphere changes back to a 3-sphere. Therefore the spatial
curvature is positive whereas the curvature of the spacetime is negative,
i.e. one has a negative curvature along the time-like component of
the curvature. We will show this fact now. Let 
\[
\tilde{W}=W_{1}\cup W_{2}\cup W_{2}\cup\ldots
\]
be the spacetime for this phase. Then using the Mostow rigidity (for
the 4-manifold $\tilde{W}$), we have a \emph{constant}
negative scalar curvature
\begin{equation}
R_{\tilde{W}}=-\Lambda<0\label{eq:exotic-hyperbolic}
\end{equation}
for the 4-manifold $\tilde{W}$. The spatial component (the 3-sphere)
is assumed to carry a metric of constant positive curvature
\begin{equation}
^{(3)}R=\frac{1}{r^{2}}\label{eq:spatial-curvature}
\end{equation}
with radius $r$. For the whole 4-manifold $\tilde{W}$ we have the
topology of $S^{3}\times[0,1)$ and we choose a product metric of
the form
\[
ds^{2}=dt^{2}-R^{2}\cdot h_{ik}dx^{i}dx^{k}
\]
with the scaling factor $R=R(t)$ and the homogenuous metric $h_{ik}$
of the 3-sphere (of unit radius). Then one obtains for the Ricci tensor
\begin{eqnarray*}
R_{00} & = & -3\frac{\ddot{R}}{R}\\
R_{ii} & = & \frac{\ddot{R}}{R}+2\frac{\dot{R}^{2}}{R^{2}}+\frac{2}{R^{2}}\qquad i=1,2,3
\end{eqnarray*}
(the dot is the time derivative) leading to a negative scalar curvature
\[
R=-\frac{6}{R^{2}}\left(R\ddot{R}+\dot{R}^{2}+1\right)
\]
(in agreement with our setting above). The spatial componentes of
the Ricci tensor are positive for $\ddot{R}\geq0$ and negative along
the time-like component. Secondly, the spatial scalar curvature (\ref{eq:spatial-curvature})
is also positive. Then according to (\ref{eq:exotic-hyperbolic})
we obtain
\[
R\ddot{R}+\dot{R}^{2}+1=\frac{\Lambda}{6}R^{2}
\]
with an accelerated expansion. Furthermore we also shown the claim
above. We call $\Lambda$ the cosmological constant. This constant can be determined by the expansion rate (\ref{eq:expansion-rate})
of the inflation. We assumed a Planck-size cosmos at the big bang
which grows to a cosmos of size
\[
L=L_{P}\cdot\exp\left(\vartheta\right)
\]
and curvature
\[
\frac{1}{L_{P}^{2}}\cdot\exp(2\cdot\vartheta)\:.
\]
Then the Mostow rigidity of the homology 3-sphere $\Sigma$ implies
a constant curvature which determines by the same argument (now for
the 4-manifold) the curvature $R_{\tilde{W}}$ to 
\[
R_{\tilde{W}}=-\Lambda=-\frac{1}{L_{P}^{2}}\cdot\exp(-2\cdot\vartheta)\:.
\]
Thus we obtain an exponential small expression for the cosmological
constant! For the homology 3-sphere $\Sigma(8_{10})$ above, we obtain
\[
\Lambda\cdot L_{P}^{2}=\exp\left(\frac{3\cdot vol(\Sigma)}{CS(\Sigma)}\right)\approx\exp(-166.262..)\approx6.2\cdot10^{-73}
\]
which is not small enough to explain the current value of the cosmological
constant. But there is a lot of freedom to construct a hyperbolic
homology 3-sphere from a knot. In particular, the closing of the knot
complement by using a so-called cusp is very important. Therefore,
for a $+3$ Dehn-surgery with a special cusp (generating a geodesic
of minimal length $0.5054$), we obtain a homology 3-sphere $\tilde{\Sigma}(8_{10})$
with 
\begin{eqnarray*}
vol(\tilde{\Sigma}(8_{10})) & = & 4.67277013...\\
CS(\tilde{\Sigma}(8_{10})) & = & 0.05095345...
\end{eqnarray*}
so that 
\[
2\cdot\vartheta\approx 275.12021...
\]
and
\[
\Lambda\cdot L_{P}^{2}\approx3.3\cdot10^{-120}\quad.
\]
In cosmology, one usually relate the cosmological constant to the
Hubble constant $H_{0}$ and the critical density leading to the length
scale
\[
L_{c}^{2}=\frac{c^{2}}{3H_{0}^{2}}\,.
\]
The corresponding variable is denoted by $\Omega_{\Lambda}$ and we
obtain
\[
\Omega_{\Lambda}=\frac{c^{5}}{3\hbar GH_{0}^{2}}\cdot\exp\left(-\frac{3\cdot vol(\tilde{\Sigma}(8_{10}))}{CS(\tilde{\Sigma}(8_{10}))}\right)
\]
in units of the critical density and using the Planck length $L_{P}=\sqrt{\frac{\hbar G}{c^{3}}}$.
By using the measured value for the Hubble constant
(Planck sattelite)
\[
H_{0}=68\,\frac{km}{s\cdot Mpc}
\]
we are able to calculate the dark energy density (as expression for
the cosmological constant)
\[
\Omega_{\Lambda}=0.513
\]
and we obtain only the rough order of the constant (in contrast to
the current value $(\Omega_{\Lambda})_{measure}=0.69$ measured by
the Planck sattelite). 

\section{Conclusion}
The paper discussed a simple model of a cosmic evolution starting
with a quantum state (represented by a wildly embedded 3-sphere) which
changed to a classical state (tame embedded homology 3-sphere $\Sigma$).
In the model we are able to calculate the decoherence time of the
quantum state. Furthermore, we obtain a small cosmological constant
with a value which is in qualitative agreement with the current measured
value. The constant value of the cosmological constant can be geometrically
understood by Mostow rigidity. We speculate that the correct value
of the cosmological should be obtained by a more realistic model including
matter coupling. 

\appendix

\section{Wild and Tame embeddings}

We call a map $f:N\to M$ between two topological manifolds an embedding
if $N$ and $f(N)\subset M$ are homeomorphic to each other. From
the differential-topological point of view, an embedding is a map
$f:N\to M$ with injective differential on each point (an immersion)
and $N$ is diffeomorphic to $f(N)\subset M$. An embedding $i:N\hookrightarrow M$
is \emph{tame} if $i(N)$ is represented by a finite polyhedron homeomorphic
to $N$. Otherwise we call the embedding \emph{wild}. There are famous
wild embeddings like Alexanders horned sphere \cite{Alex:24} or Antoine's
necklace. In physics one uses mostly tame embeddings but as Cannon
mentioned in his overview \cite{Can:78}, one needs wild embeddings
to understand the tame one. As shown by us \cite{AsselmeyerKrol2009},
wild embeddings are needed to understand exotic smoothness.

\section*{References}

%\bibliographystyle{unsrt}
%\addcontentsline{toc}{section}{\refname}
%\bibliography{measurements,foliation-gerbes,knots,diffbib,inflation}

\end{document}